
\input harvmac
\Title{\vbox{\baselineskip14pt\hbox{IMSc--92/30}\hbox{hepth@xxx/9207109}}}
{Chiral Rings and Physical States in $c<1$ String Theory}

\centerline{Suresh Govindarajan, T. Jayaraman and Varghese
John\foot{email:suresh, jayaram, john@imsc.ernet.in}}
\centerline{The Institute of Mathematical Sciences}
\centerline{C.I.T. Campus, Taramani}
\centerline{Madras 600 113, INDIA}
\bigskip
\bigskip

We  show how
the double cohomology of the String and Felder BRST charges naturally
leads to the ring structure of $c<1$ strings.
The chiral ring is  a ring of polynomials in two variables modulo
an equivalence relation of the form $x^p \simeq y^{p+1}$ for the (p+1, p)
model. We also
study the states corresponding to the edges of the conformal grid whose
inclusion is crucial for the closure of the ring. We introduce
candidate operators that correspond to the observables of the matrix models.
Their existence is motivated by the relation of one of the screening
operators of the minimal model to the zero momentum dilaton.

\Date{July 1992}
\lref\SAR{H. Kanno and H. Sarmadi, Talk presented by H. Sarmadi at the
Summer Workshop on Strings at ICTP, Trieste(July 1992).}
\lref\SARa{H. Kanno and H. Sarmadi, ``BRST cohomology ring in 2D gravity
coupled to minimal models,'' ICTP preprint IC/92/150 =
hepth@xxx/9207078}
\lref\DOY{M. Doyle, ``Dilaton Contact Terms in the Bosonic and Heterotic
Strings,'' Princeton preprint PUPT-1296(1992) =
hepth@xxx/9201076.}
\lref\BT{R. Bott and L. W. Tu, {\it Differential Forms in Algebraic
Topology}, Springer-Verlag(1982).}
\lref\KACH{S. Kachru, ``Quantum Rings and Recursion Relations in 2D Quantum
Gravity,'' Princeton Preprint PUPT-1305 = hepth@xxx/9201072.}
\lref\ITOH{K. Itoh, ``SL(2,R) current algebra and spectrum in two
dimensional gravity,'' Texas A\&M preprint CTP-TAMU-42/91(1991).}
\lref\AGG{L. Alvarez-Gaum\'e and C. Gomez, ``Topics in Liouville Theory,''
Lectures at Trieste Spring School, CERN preprint CERN-TH.6175/91, (1991).}
\lref\GRO{U. H. Danielsson and D. Gross, \npb{\bf 366}(1991), 3.}
\lref\FKa{P. Di Francesco and D. Kustasov, \npb{\bf 342}(1990), 589.}
\lref\FKb{P. Di Francesco and D. Kutasov, ``World Sheet and Space time
Physics in two dimensional (super) string theory,'' Princeton preprint
PUPT-1276(1991).}
\lref\DOT{V. Dotsenko, ``Three Point Correlation Functions of the Minimal
Conformal Theories coupled to 2D gravity,'' \mpl{\bf 6}(1991), 3601.}
\lref\GF{G. Felder, \npb317 (1989), 215.}
\lref\LZ{B. Lian and G. Zuckerman, \plb {\bf 254}(1991), 417; ``Semi-infinite
homology and 2D Gravity I,'' \cmp{\bf 145}(1992), 561.}
\lref\IMM{C. Imbimbo, S. Mahapatra and S. Mukhi,``Construction of Physical
States of Non-trivial ghost number in $c<1$ String Theory,'' \npb375(1992),
399.}
\lref\GLI{M. Goulian and B. Li, \prl{\bf 66}(1991), 2051.}
\lref\BMP{P. Bouwknegt, J. McCarthy and K. Pilch, ``BRST analysis of physical
states for 2d gravity coupled to $c<1$ matter,'' \cmp145(1992), 541.}
\lref\BMPa{P. Bouwknegt, J. McCarthy and K. Pilch, ``Fock Space Resolutions
of the Virasoro highest weight modules with $c\leq1$,'' Lett. Math. Phys.
23(1991), 3601.}
\lref\KIT{Y. Kitazawa, \plb {\bf 265}(1991), 262.}
\lref\KITa{Y. Kitazawa, ``Puncture Equation in $c=1$ Liouville
gravity,'' TIT preprint TIT(1991).}
\lref\KM{M. Kato and S. Matsuda in {\it Advanced Studies in Pure Mathematics},
Vol. 16, ed. H. Morikawa(1988), 205.}
\lref\POL{A. M. Polyakov, \mpl{\bf 6}(1991), 635-644.}
\lref\FF{B. Feigin and D. Fuchs, ``Representations of the Virasoro algebra,''
in {\it Seminar on Supermanifolds} No.5, ed. D. Leites(1988), Univ. of
Stockholm Report No. 25.}
\lref\JD{J. Distler, \npb{\bf 342}(1990), 523.}
\lref\DN{J. Distler and P. Nelson, ``New discrete states of strings
near a black hole,'' Penn preprint UPR-0462T(1991).}
\lref\FUT{S. Govindarajan, T. Jayaraman, V. John, and P. Majumdar,
work in progress.}
\lref\ESG{E. S. Gardner, unpublished.}
\lref\KAL{S. Kalyana Rama, ``New special operators in W-gravity theories,''
Tata preprint TIFR/TH/91-41(1991).}
\lref\SEI{N. Seiberg,``Notes on Quantum Liouville Theory and Quantum Gravity,
,'' Prog. of Theo. Phys., {\bf 102}(1990), 319.}
\lref\POLC{J.Polchinski, ``Remarks on the Liouville field theory,'' Texas
preprint UTTG-19-90, in Proceedings of Strings '90.}
\lref\BER{M.Bershadsky and I.Klebanov, \prl{\bf 65}(1990), 3088;
\npb{\bf 360}(1991), 559.}
\lref\KAED{K. Aoki and E. D'Hoker, ``On the \lio\ approach to correlation
functions for 2-D quantum gravity,'' UCLA preprint UCLA/91/TEP/32(1991).}
\lref\GJJM{S. Govindarajan, T. Jayaraman, V. John and P. Majumdar,
``States of non-zero ghost number in $c<1$ matter coupled to 2d
gravity,'' Mod. Phys. Lett. {\bf A7}, (1992)1063-1077.}
\lref\DOTa{Vl. M. Dotsenko, `` Remarks on the physical states and the
chiral algebra of 2d gravity coupled to $c\leq1$ matter,''
PAR-LPTHE 92-4(1992) = hepth@xxx/9201077;  ``The operator algebra
of the discrete state operators in 2D gravity with
non-vanishing cosmological constant,''
CERN-TH.6502/92 =
PAR-LPTHE 92-17.}
\lref\EW{E. Witten, ``Ground Ring of two dimensional string theory,''
\npb{\bf 373}(1992) 187.}
\lref\KMS{D. Kutasov, E. Martinec and N. Seiberg, ``Ground Rings and
their modules in 2d gravity with $c\leq1$ matter,'' \plb{\bf 276}(1992), 437.}
\lref\KLEB{I. R. Klebanov, ``Ward Identities in 2d gravity,''
Mod. Phys. Lett. {\bf A7}, (1992)723,}
\lref\CDK{N. Chair, V. Dobrev and H. Kanno, ``$SO(2,{\bf C})$
invariant ring structure of BRST cohomology and singular vectors in 2d
gravity with $c<1$ matter,'' \plb{\bf 283}(1992), 194.}
\lref\EWBZ{E. Witten and B. Zweibach, ``Algebraic structures and
Differential Geometry in 2d string theory,'' \npb{\bf 377}(1992), 55.}
\lref\JDTG{J. Distler, ``2-d quantum gravity, topological field theory,
and the multicritical matrix models,'' Nucl. Phys. {\bf B342} (1990) 523.}%
\lref\TCCT{J. Distler and P. Nelson, ``Topological couplings and
contact terms in 2d field theory,''
Commun. Math. Phys. {\bf 138} (1991) 273.}%
\lref\dil {J. Distler and P. Nelson, ``The dilaton
equation in semirigid string theory,''  PUPT-1232 = UPR0428T (1991),
Nucl. Phys. {\bf B}, in press.}
\lref\GJJ {S.Govindarajan, T. Jayaraman and V. John, IMSc preprint
IMSc--92/31, to appear}

\def\plb{Phys. Lett. {\bf B}}
\def\prl{Phys. Rev. Lett.}
\def\mpl{Mod. Phys. Lett. {\bf A}}
\def\npb{Nuclear Phys. {\bf B}}
\def\cmp{Comm. Math. Phys.}
\def\ket#1{| #1 \rangle}
\def\melt#1#2#3{\langle #1 \mid #2 \mid #3 \rangle}
\def\lfr#1#2{\textstyle{#1 \over #2 }}
\def\al{{\alpha}}
\def\pa{{\partial}}

\def\lio{Liouville}
\def\vm#1#2{\ket{v_{{#1},{#2}}}_M}
\def\um#1#2{\ket{u_{{#1},{#2}}}_M}
\def\wm#1#2{\ket{w_{{#1},{#2}}}_M}
\def\vl#1#2{\ket{v_{{#1},{#2}}}_L}
\def\gh{c_1\ket0_{gh}}

\newsec{Introduction}
The continuum formulation of $c<1$ models coupled to gravity continues
to be much less developed in comparison to the progress that was made
using the matrix model approach. Physical states,
physical operators and their correlation functions are not yet fully
understood, especially if we demand that they are in one-to-one
correspondence with the results of the matrix model. Clearly, the source
of the problem lies in the fact that in the continuum formulation, the
Liouville field is considered to be a scalar field and the full theory is then
treated by the machinery of standard string theory. What in essence is
necessary therefore, is some kind of consistent truncation of this theory to
reproduce the results of the matrix models.

These problems were already visible from the work of Lian and Zuckerman \LZ
in their construction of the BRST invariant states of the minimal models
coupled to gravity. The whole set of positive ghost number states
had scaling dimensions that did not match those that came
from the matrix model. On the other hand, even the negative ghost number
states needed to be supplemented by those that came from the boundary of
the Kac table. It seemed that a correlation function could not contain
operators that matched those of the matrix model while at the same time
conserving ghost number so that it did not vanish\AGG. A further problem
with these states was the fact that their construction was not
immediately obvious.  Some progress was also made in correlation functions
when the work of Goulian and Li\GLI, Dotsenko \DOT and Kitazawa \KIT
showed how three-point functions could be calculated using the Coulomb
gas formalism and that they were the same as those of
the matrix model. It was also found possible to do the calculations for
states both inside and outside the Kac table\DOT\KIT. The calculation of higher
point functions was however blocked by the necessity of analytic
continuation in the number of screening operators to negative or
fractional values.

Some of these problems began to be addressed by the use of the full
power of the Coulomb
gas method to represent the matter part of the operators. Apart from
being useful to study the BRST cohomology itself\LZ, we showed that
this method was
helpful in indicating how these states could be explicitly written down
\GJJM. This was particularly useful for states of ghost number different
from $\pm 1$ \foot{In our
convention the state $\ket0_M\otimes\ket0_L\otimes\gh$ is defined to
have zero ghost number with $c(b)$ having ghost number $+1(-1)$.
Further, in going from states to operators, one increments the ghost
number by 1.}
(these were constructed explicitly earlier by  different
methods \IMM \BER ).
Another advantage of the method was that it provided an alternate
representation of these BRST invariant operators in terms of pure vertex
operators that were easier to work with. Indeed it was shown that ghost
number conserving correlation functions of the Lian-Zuckerman operators
could be mapped into those involving only pure vertex operators(DK
states).

In this paper we begin by reviewing the Coulomb gas construction of the
BRST invariant states and operators using the double cohomology of the
Felder BRST and the string BRST charges. We extend our considerations to
the states that come from the boundary of the Kac table and show how
they naturally appear in the Coulomb gas construction. The inclusion of
both primary fields and their duals leads to the appearance of ghost
number zero operators that provide the chiral ring structure for these
models analogous to those first studied by Witten \EW \EWBZ in the context
of $c=1$
theories coupled to gravity. We show some explicit construction of these
operators and provide a general cohomology argument for their existence.
The appearance of this structure is extremely useful as it provides a
means to compute quite general correlation functions of vertex operators
for states both inside and outside the Kac table. In fact, such
computations are crucial in showing that we can compute correlation
functions with generalized screening operators which enables one to  avoid
the problem of analytic continuation in the number of screening
operators\GJJ. This opens the route to computation of n-point functions
in the continuum formulation.

We finally turn to a
consideration of the true physical operators of the theory, namely the
operators that can be both ghost number conserving and have the same
scaling dimensions as the operators in the matrix model.
Of course, we are not referring here to operators that arise purely as
vertex operators as one may object that they are purely artefacts of the
Coulomb gas method. What we are interested in are operators that can equally
well be represented directly in terms of the primary and secondaries in the
matter part of the theory. Thus we are interested in physical operators
that are built out of the higher ghost number states that appear in the
Lian and Zuckerman analysis.
It appears that
we have to go beyond the considerations  of the original BRST analysis.
We propose, in analogy with the dilaton of usual string theory, a set of
operators that are BRST trivial in the holomorphic sector of
the theory and have positive ghost number and are representative of
the BRST  cohomology
of negative ghost number in the anti-holomorphic sector. We
motivate the construction using in  particular
the relation of the screening operator to the zero momentum dilaton. We also
discuss how these operators may be directly related to the matrix model
observables.
We relegate a mathematical argument
to the appendix.

\newsec{BRST Invariant States in $c<1$}
In the continuum formulation of $c<1$ matter coupled to two
dimensional gravity, the cohomology analysis of Lian and Zuckerman\LZ and
subsequently by Bouwknegt, McCarthy and Pilch\BMP has shown that the
physical states occur at non-trivial ghost number
-- one for
each null state over the primaries. The Liouville dressing of these
states of non-zero ghost number is the same as the dressing of the
corresponding null states. We refer to these states(of non-trivial
ghost number) as LZ states.

States outside the
minimal conformal grid(Kac table), when provided with a suitable
Liouville dressing will be called
DK states.\foot{DK states
are sometimes referred to as tachyon states.} DK states have $0$ ghost
number in our convention. In this section, we will explain how DK states
and ring elements(ghost number -1 states) provide an equally good
representation of physical states. However,
unlike LZ states, these states are easy to write down
and one can actually attempt calculations involving these operators.
In an
earlier paper\GJJM, we had clarified how DK states are actually
related to  LZ states by means of
descent equations which follow from a double cohomology analysis.
We shall now briefly summarise those results and then proceed to explain
how ring elements are also related to LZ states via descent.

\subsec{Double Cohomology}
We consider two scalars $X$(for matter) and $\phi$(for the Liouville
mode) with background charges $\al_0$
and $\beta_0$ respectively at infinity. The corresponding
energy-momentum tensors
are given by
\eqn\estress{\eqalign{
T^M &= -\lfr14\pa X\pa X + i \al_0 \pa^2 X\quad,\cr
T^L &= -\lfr14\pa\phi\pa\phi + i \beta_0 \pa^2\phi\quad,
}}
with central charges $c_M=1-24\al_0^2$ and $c_L=1-24\beta_0^2$. For
the
$(p,p+1)$ unitary minimal models
$$
\al_0^2=\lfr1{4p(p+1)}~ {\rm and}~\beta_0^2=
-\lfr{(2p+1)^2}{4p(p+1)}\quad.
$$
 The vertex operators $e^{i\al X}$ and
$e^{i\beta\phi}$ have conformal weights $\al(\al-2\al_0)$ and
$\beta(\beta-2\beta_0)$ respectively. The usual screening charges for matter
are
$$
\al_+=\sqrt{\lfr{p+1}p}~{\rm and}~\al_-=-\sqrt{\lfr p{p+1}}\quad.
$$
The screening charges for the \lio\ sector are given by
$$
\beta_+=i\al_+~{\rm and}~\beta_-=-i\al_-\quad.
$$
Following Felder\GF, we consider the complex of Fock spaces
(hereafter referred to as the Felder complex)
in the matter sector given by $\bigoplus ~F_{m',\pm m+2np}$, where $F_{m',m}$
is the Fock space built over the primary associated with the vertex operator
$e^{i\al_{m',m}X}$. Here
$$
\al_{m',m}=\lfr{1-m'}2\al_- + \lfr{1-m}2\al_+\quad.
$$
We will also need the dual Fock space obtained by $F_{-m',-m}$.
There exists an identity under the change of label given by $(m',m)\rightarrow
(m'+p+1, m+p)$ for the matter sector and $(m',m)\rightarrow(m'+p+1,m-p)$ in the
\lio\ sector. These two seemingly distinct labels refer to the same
vertex operator. This identity will prove to be useful later.
There is one such Felder complex for every $m',m$ restricted to the range
$1\leq m'\leq p$ and $1\leq m \leq (p-1)$. The screening operators
$Q_m^+ = \int \prod_{i=1}^m dz_i e^{i\al_+X(z_i)}$ and similarly $Q_{p-m}^+$
act on these Fock spaces. The irreducible Virasoro module $L(m',m)$
(for a given $c_m$ labelled by $p$) is given by $Ker~Q_m^+/Im~Q_{p-m}^+$ on
this
complex. We shall refer to the screening operators loosely as $Q_F$ except
when necessary.
 We also have the Fock spaces of the \lio\ and ghost sectors denoted
by $\CF(\beta)$ and $\CF(gh)$ respectively. The string BRST operator $Q_B$
given by
\eqn\ebrst{
Q_B = \oint :c(z) (T^M(z) + T^L(z) + \lfr12 T_{gh}(z)):
}
acts on the tensor product $\CF(\al)\otimes\CF(\beta)\otimes\CF(gh)$.

Hence, in the the Coulomb gas realisation of minimal models coupled to gravity
one has to work with the double complex associated with the two BRST
operators.

\subsec{Descent Equations}
The analysis of the double cohomology of the string and Felder BRST
\LZ\BMP\GJJM provides a simple relation between LZ states of non-trivial
ghost number and DK states. This is obtained by means of descent equations
as was explicitly shown in \GJJM.
The existence of descent is seen from the isomorphism of cohomology
classes\BMP,
\eqn\ecoha{\eqalign{
H^{(n)}_{rel}(
H^{(0)}(\CF(\al)_M&\otimes\CF(\beta)_L\otimes\CF_{gh},~Q_F),~Q_B)\cr
&\simeq
H^{(n)}(
H^{(0)}_{rel}(\CF(\al)_M\otimes\CF(\beta)_L\otimes\CF_{gh},~Q_B),~Q_F)\quad.
}}
Since Felder\GF, has shown that $H^0(\CF(\al_{m',m}))\simeq L(m',m)$,
the LHS of \ecoha are the LZ states and the RHS are the DK states.
This implies that the LZ and DK states are isomorphic to each other
and hence the existence of descent.

The descent equations are
\eqn\edesca{\eqalign{
Q_B {\ket{LZ}}^{-n} &= Q_F{\ket{I_1}}^{-n+1}\cr
&\vdots\cr
Q_B {\ket{I_{n-1}}}^{-1} &= (-)^{(n-1)} Q_F{\ket{DK}}^{0}
}}
where we have denoted the ghost number of the state in the
superscript. One can always use \edesca to construct LZ states by
``solving'' the equations as has been demonstrated in \GJJM.
For the generic ghost number $-1$ LZ state, we have
\eqn\edescd{
Q_B {\ket{LZ}}^{-1} = Q_F{\ket{DK}}^{0} = \ket{u}_M\otimes\ket{v}_L\otimes\gh
}
where $\ket{u}_M$ is the non-vanishing matter null. One important point is
that the LZ states thus obtained come with a precise choice of matter momenta.
For the $-1$ case the matter momenta is such that there exists a non-vanishing
null at the required level.

Consider the following example, already dealt
with in \GJJM for the case of $c_M=0$.
The descent equation for a ghost number $-2$ LZ state  is
\eqn\edescb{\eqalign{
    Q_B \ket{LZ}& = Q_F \ket{I}\quad,\cr
    Q_B \ket{I} &= -Q_F \ket{DK}\quad.
}}
Figure 1 shows the location of the matter part of these states in the
Felder complex.
Consider the DK state obtained from dressing $\vm25$. ``Solving'' \edescb\
and after tedious but straightforward algebra, we obtain
\eqn\elzb{\eqalign{ \ket{DK} &=\vm25\otimes\vl{-2}5\otimes\gh\quad,\cr
 \ket I &= \CL^b_{4} \vm23\otimes\vl{-2}5\otimes\gh\quad,\cr
 \ket{LZ} &= \CL^{2b}_{5}\vm21\otimes\vl{-2}5\otimes\gh\quad,
}}
where
\eqn\eec{\eqalign{
\CL^b_{4} &= \lfr{94}3 b_{-4} + b_{-3}(\lfr{61}3 L^L_{-1} +3L^M_{-1})\cr
           &+ b_{-2}(4L^L_{-2}- 4L^M_{-2} +\lfr{20}3L^{L~2}_{-1})\cr
           &+b_{-1}(-3L^L_{-3} -\lfr{41}3L^M_{-3}-\lfr{20}3L^L_{-1}L^M_{-2}\cr
   &+\lfr{20}3L^M_{-2}L^M_{-1} + L^{L~3}_{-1} -L^{L~2}_{-1}L^M_{-1}
   + L^{L}_{-1}L_{-1}^{M~2}-L^{M~3}_{-1})\quad,
}}
and\foot{We correct a typographical error in the expression for
$\CL^{2b}_5$ given in \GJJM}
\eqn\eed{\eqalign{
\CL^{2b}_{5} &= -\lfr43b_{-4}b_{-1} + 4b_{-3}b_{-2}
+b_{-3}b_{-1}(\lfr23L^L_{-1}-15L^M_{-1})\cr
              &+ b_{-2}b_{-1}(-4 L^L_{-2} -\lfr23L^{L~2}_{-1}
-6L^L_{-1}L^M_{-1} + 6 L^{M~2}_{-1})\quad.
}}
These expressions are unique upto $Q_B$ exact pieces that are also $Q_F$
closed. One thus obtains the LZ state with matter momentum labelled by $(2,1)$
as seen in \elzb.

The construction outlined above is tied down to a particular choice of
matter momentum in the LZ state. Since there exists an equivalent LZ state
with dual matter momentum,
one would like to know what happens to descent if
we had taken the dual $(2 \alpha_0 - \alpha)$ in the matter momentum of the
LZ state.
The simple case of a ghost number $-1$ state illustrates the
point. Let $\ket{\widetilde{LZ}}$ represent the state with matter momenta dual
to that obtained from the LZ state in \edescd. Now the operation of $Q_B$
on $\ket{\widetilde{LZ}}$ again gives the matter null. But, this is a
vanishing null and hence descent ends at this point.
Thus the ghost number $-1$ state itself is the other representative at
the end point of the descent instead of a DK state.

This situation generalises to the ghost number $-n$ case, with the descent
ending at a ghost number $-1$ state instead of a ghost number
zero state (viz. a DK state) on flipping the matter momenta of the LZ state
to its dual. The flip of matter momenta to its dual interchanges
non-vanishing nulls with the vanishing nulls. Thus a descent from the LZ state
would now give an intermediate state with different matter momenta and
level different from the original one.
For instance, in the contruction of a state of ghost number $-2$ given earlier
in \edescb(figure 2 depicts the descent on the dual Felder complex). Now we
have
\eqn\eringa{\eqalign{
\ket{\widetilde{LZ}}^{-2} &= \CL^{2b}_{5}\vm11\otimes\vl{-2}5\otimes\gh\quad\cr
Q_B \ket{\widetilde{LZ}}^{-2}&=
                         \CL^b_3\CL_2\vm11\otimes\vl{-2}5\otimes\gh\quad\cr
&=-4 \CL^b_3\um11\otimes\vl{-2}5\otimes\gh\quad\cr
&=Q_F\ket{R}^{-1}
}}
where
$$
\eqalign{
\CL^b_3 &=  b_{-3} -b_{-2} L_{-1}^{M} + b_{-1} (L_{-2}^{L} + \lfr16(L_{-1}^{L~2
} -L^{L}_{-1}L_{-1}^{M} +L_{-1}^{M~2})) \quad,\cr
\CL_2 &= (L^M_{-2} - \lfr32 {L^M_{-1}}^2)\quad. \cr
}
$$
In the second step of the
descent in \edescb, we had  the non-vanishing null over $\vm23$  appearing in
the
construction. Here we obtain the vanishing third level null over $\vm13$.
and thus $Q_B\ket{R} = 0$. Thus the descent terminates one step earlier.

{}From a constructionist viewpoint, one can now see how such a process can
occur for states of arbitrary ghost number $-n$. An examination
of the levels of the appropriate null vectors in the
Felder diagram is in agreement with a descent terminating earlier.
A more general and rigourous argument using dimensionalities of
cohomology classes in the double cohomology of $Q_B$ and $Q_F$ shows that
indeed such ghost number -1 states exist. We present the details of the
argument in the Appendix. Hence there are two possible endpoints for descent:
states at zero ghost number -- DK states and states
at ghost number $-1$. The latter are precisely the ring elements for $c<1$.
We shall discuss them in more detail in the following section.

So far the discussion has been restricted to the negative ghost number
sector. The positive ghost number states are partners to the negative ghost
number states in the sense that the norm on the sphere is obtained as
\eqn\enorm{
\phantom{a}^{+n}\melt{LZ}{c_0}{LZ}^{-n}
}
Given an LZ state of ghost number $-n$, it is now trivial to construct a
$Q_B$ closed state of ghost number $+n$ which has a non-zero norm with
the given LZ state. This new state of ghost number $+n$ is
\eqn\epos{
|LZ\rangle^{+n}=M^n |\widehat{LZ}\rangle^{-n}\quad,
}
where $M=\{Q_B,c_0\}$ and $ |\widehat{LZ}\rangle^{-n}$ is the LZ state
with the matter and Liouville momenta flipped to their duals. The state
given in \epos\ is obviously not exact and hence a good element of the
cohomology. The Liouville dressing is $\beta>\beta_0$ as
given by the analysis of Lian and Zuckerman. It can also be shown that
the LZ states of positive ghost number thus obtained are equivalent to
those obtained by the construction described in \GJJM\  upto exact
pieces.
\subsec{Edge States}
       The edge states are matter states corresponding to the boundary
of the Kac table i.e, $(m',0)$ and $(0,m)$. In the minimal models before
coupling to gravity one can use the fusion rules to show that the edge
states decouple (along with the states outside the Kac table), and the
states inside the Kac table form a closed algebra in the usual OPE
sense.

     However, on coupling to gravity the minimal model fusion rules
seem to be erased \DOT \KIT. Hence we find that the edges (and the states
outside the Kac Table) do not decouple by the standard arguments. In fact
at the level of the three point functions it was shown by Kitazawa \KIT\
that even though the matter contribution to the three point function
is vanishing the Liouville contribution is infinite so that the full
three point function is finite, upto singular gamma functions
corresponding to the leg  factors of the external states. We emphasise that
this non-decoupling of the edge states is not dependent on the Coulomb gas
formalism. The zeros of the OPE coefficients are cancelled by the infinities
of the Liouville and these OPE coefficients carry intrinsic meaning independent
of the Coulomb gas method. Indeed the order of the zero, for instance can
easily obtained from the monodromy coefficients of the minimal model
correlation functions.

We now look at the BRST analysis of these states on coupling to gravity.
This has been previously studied in \LZ \BMP . We recollect these
results briefly in this section and  then indicate how descent
equations can be used to construct the corresponding LZ states.

   The Fock space resolutions of the edge states have been discussed
in \BMPa\ and they are of the form(see figure 3)
$$ 0\buildrel Q_F \over \longrightarrow {\cal F}_{0 ,m- (k+2)p}
\buildrel Q_F \over \longrightarrow {\cal F}_{0, m+ kp}
\buildrel Q_F \over \longrightarrow  0 $$
and in the dual space(see figure 4)
$$ 0\buildrel Q_F \over \longrightarrow  {\cal F}_{0, -m- kp}
\buildrel Q_F \over \longrightarrow   {\cal F}_{0,-m+(k+2)p}
\buildrel Q_F \over \longrightarrow   0 $$

where $Q_F =(Q_-)^{(k+1)(p+1)}$.

   Notice that since the Felder complex associated with the edge states
has only finite number of Fock spaces,
there are only LZ states of ghost number $(\pm1)$. The matter part of the
LZ states belong to the Fock space labelled $(0, m+kp)$ and its dual
respectively.
Now we use the descent
equations which arise from the double cohomology analysis of the Felder
and string BRST analysis (as was done for the states in the interior of
the Kac table in  \GJJM ).
For the LZ state of
positive ghost number we have(see figure 4)
\eqn\desa{\eqalign{
\ket{DK}&= Q_F {\ket{v_{0,-m-kp}}}_M\otimes\vl0{-m+(k+2)p}\otimes\gh\quad,\cr
\ket{LZ}^{+1}&=Q_B{\ket{v_{0,-m-kp}}}_M\otimes\vl0{-m+(k+2)p}\otimes\gh\quad,\cr
}}

The descent equation for the ghost number $-1$ is(see figure 3)
\eqn\desb{
Q_B {\ket{LZ}}^{-1}_{(0,m+kp)} = Q_F{\ket{u'_1}}_{(0,m-(k+2)}}
from which we obtain the LZ state with ghost number (-1). As in the case of
non-edge states, on taking the matter dual in the LZ state of ghost number
$-1$, we obtain the ring elements corresponding to the edge states. These
have matter momenta labelled by $(0,-m-kp)$.

  A similar construction can be done for the $(m',0)$ edges using
$Q_F=(Q_+)^{(k+1)p}$. Unlike the non-edge states, the descent seems tied
down to a particular choice of resolution, i.e., descents involving $(0,m)$
edges use only
$Q_-$ and $(m',0)$ use only $Q_+$. Of course, this does not matter for the case
of the ring elements as in this case there is no descent.
Notice that in the construction of the LZ states of the edges we
used Felder BRST operators of the form $Q_F= (Q_-)^{(k+1)(p+1)}$
unlike in the construction for nonedge states where we always used
Felder BRST operators of the form $Q_F= (Q_+)^{m}$ with $m<p$.

 On comparing the scaling exponents of matrix models and
topological minimal models coupled to gravity, only the $(0,m)$ edge states
are required to complete the series of exponents. For the $(p+1,p)$ models,
the Liouville momenta are parametrised by the series
\eqn\elio{
{p(n-2)+\al}\over{2\sqrt{p(p+1)}}
}
where $n\geq0$ and $\al=0,1,\ldots,(p-2)$\BER. The edge states of type
$(m',0)$
are outside this set and correspond to $\al=(p-1)$.
 Thus, the two sets of edge
states, although similar at first sight
actually behave differently. However, it is not easy to see how the ``wrong''
edge states -- $(m',0)$ actually decouple in OPEs and correlation functions.

\newsec{Chiral rings in $c<1$}

The general construction which argues for the presence of ghost number
$-1$ states in the Coulomb gas method allows us to develop a ring
structure in analogy with the work of Witten for $c=1$. Following the
suggestion of Kutasov, Martinec, and Seiberg\KMS, we define the two
operators that generate the ring structure,
\eqn\egena{\eqalign{
x&=\CR_{1,2}=(b_{-2}c_1 + t(L_{-1}^{L} - L_{-1}^{M})
e^{i\al_{1,2}X}e^{i\beta_{1,2}\phi}
\cr
y&=\CR_{2,1}(b_{-2}c_1 + \lfr1t(L_{-1}^{L} - L_{-1}^{M})
e^{i\al_{2,1}X}e^{i\beta_{2,1}\phi}
}}
i.e., $x$ and $y$ are the LZ$^{-1}$ states with matter momenta
labelled by $\al_{1,2}$ and $\al_{2,1}$ respectively. Before we proceed
further we would like to point out that a target space  boost of the
$(X, \phi)$ system which transforms the $c=1$ theory to the appropriate
$c<1$ theory , would in fact transform the generators $x$ and $y$ of
Witten to precisely the ones we have written above. This fact is quite useful
and we return to it later on.

For simplicity, let us consider the case of $c=0$. In this case, $y$ is a
physical operator in  the theory while $x$ is not. The full set of Liouville
momenta for this case(including the edge states of type $(0,m)$) fit the
series $\lfr{(n-2)}{\sqrt6}$. If we examine all
the Liouville momenta that arise when we consider the powers $y^n$,
we find that they precisely fit the set that appear for all LZ states of
negative non-zero ghost number. Hence it appears that $y^n$ gives all
the ghost number zero operators whose construction we described in the
last section provided we work in the $Q_-$ resolution.
One can also check the level of the oscillator and ghost part of the
state $y^n$ . It is easy to show by examining the vertex operator
momenta for both the matter and Liouville part that the level is
$(n+1)$. A straightforward computation of the OPE of y and y in fact
shows that $y^2$ is a ghost number -1 state of the type described in the
previous section. One can expect therefore all $y^n$ to behave
consistently in this fashion.

For the case of $c=0$, $x$ as noted earlier is not a physical state.
In fact, $x$ is a state from the edge of the Kac table, however this
being the edge that does not appear in the set of operators that match
the Liouville momenta that come from the matrix model scaling
dimensions. However, if we compute $x^2$ by the
usual OPE we find that it is a physical state. We can now check the
Liouville momenta of the elements of the ring given by $x^{2n}$,
$yx^{2n}$, and $y^2x^{2n}$
span the entire set of
allowed Liouville momenta of the negative ghost number states of $c=0$
coupled to matter.  It thus appears that we have two sets of ghost
number zero operators that describe the same set of states, even though
the general cohomology argument assures us that there is only one ring
element for a given Liouville momenta.

How are the two sets of ring
elements related? The point is clarified by examining the two ring elements
$x^2$ and $y^3$ in detail. Both these operators have the same Liouville
momenta. The matter vertex operator momentum for the two operators are
$(1,3)$ and $(1,-1)$ respectively. The level of the two operators are 3
and 4 respectively. It is easy to see that if we operate with $Q_B$ on
both these states we would get zero as there are null states at these
levels. However using a  $Q_+$ descent with $x^2$ and a
$Q_-$ descent with $y^3$ we can go
back and construct the same LZ state of ghost number $-2$.
Thus we are led to an equivalence relation $x^2 \simeq y^3$. In general, all
states which under $Q_+$ or $Q_-$ descent lead to the same LZ state would
be identified with the same ring element.
Thus the chiral ring in $c=0$ matter coupled to Liouville is described
by the elements $x^{m} y^n$ modulo the relation $x^2 \simeq y^3$ and we can
have the elements $y^n$  and $xy^n$ or $x^{2n}$, $yx^{2n}$, $y^2x^{2n}$,
$x^{2n+1}$, $x^{2n+1}y$ and $x^{2n+1}y^2$. The choice is dictated by the
choice of resolution. Either set of elements forms a
ring.  To reproduce
only the allowed Liouville momenta the ``wrong'' edge states must be removed,
namely the elements $xy^n$ or $x^{2n+1},~x^{2n+1}y$ and $x^{2n+1}y^2$.

These features clearly generalise to the arbitrary $(p+1,p)$ model coupled to
gravity. Thus in the general case we have the ring generated by the elements
$x$ and $y$ and modded out by the equivalence relation
\eqn\reln{ x^p \simeq  y^{(p+1)}.}
This gives us the elements
$y^n$, $xy^n$, $x^{2}y^n, \ldots, x^{p-2}y^n$, $x^{p-1}y^n$.
The Liouville scaling dimensions of these elements run over all
the allowed Liouville scaling dimensions of the model given in \elio\ plus
the momenta
of the ``wrong'' edge states. Alternatively, we can
have the elements of the dual resolution which can be simply obtained by
using the equivalence relation \reln\ on the first set of elements.
Both these sets
of elements form a ring in themselves. Note however that unlike the case of
$c=0$, here the ``wrong'' edge states are needed for the ring to be closed
under multiplication. Hence they cannot be excluded in any obvious way.

Let us return to the case of the `unphysical' element $x$ in the case of
$c=0$.
However this state is certainly in the $Q_B$ cohomology,
except that the particular matter representation is simply not included
in the theory. Hence it would seem that another way to represent the
situation is to take the ring with elements $x^{m}y^{n}$ and mod out by
the ideal generated by $x$.

These features can clearly be generalised for the arbitrary $(p+1,p)$ model
coupled to gravity. In the general case, we have the ring with elements
$x^{m}y^{n}$ modded out by the ideal generated by the element $x^{p-1}$.
Itis easy to see that this construction gives us the elements
$y^n$, $xy^n$, $x^{2}y^n, \ldots\ldots,$ $ x^{p-2}y^n$.
Here the ``wrong'' edge  states clearly get excluded. It is
useful to see the striking similarity with the structure of the states
in topological minimal models coupled to topological gravity if we think of
the $y^n$ as coming from the gravity part and the $x$, $\ldots$, $x^{p-1}$
from the matter part.

Before we leave this  discussion of the ring structure, we would like to
return to the relation between the $c=1$ coupled to gravity
theory and the $c<1$ coupled to gravity theory that was briefly
referred to at the beginning of this section. The operators of
the two theories are related by a Lorentz transformation of the
$X$ and $\phi$ co-ordinates of the $c=1$ theory
that relates operators of that theory to those in $c<1$
coupled to gravity. This transformation has been noted by several
authors earlier (See for instance \ITOH\DOTa\CDK). The transformation of $X$
 and $\phi$  can be written down
\eqn\etrans{\eqalign{
X &\rightarrow X cosh\theta + i\phi sinh\theta \cr
\phi &\rightarrow iX( -sinh \theta) + \phi cosh \theta
}}
where $cosh \theta = \lfr{2p+1}{2\sqrt{p(p+1)}}$ and $sinh\theta =
\lfr{1}{2\sqrt{p(p+1)}}$
It is easy to see that this transformation rotates the energy-momentum
tensor of the $c=1$ model to that of the $c<1$ model with the appropriate
charge at infinity term. In the case of vertex operators, operators
with specific matter and Liouville momenta are rotated from the $c=1$
theory to the allowed matter and Liouville momenta of the $c<1$ theory.
A vertex operator of the form
\eqn\etransb{
V^{-}_{k} = e^{-ikX}e^{(-1- k)\phi}
}
is transformed to
\eqn\etransc{
V^-_{k} = exp\{\lfr1{2\sqrt{p(p+1)}}( [2(p+1)k+1]iX -[2(p+1)k +(2p+1)]\phi)\}
}

and

\eqn\etransd{
V^{+}_{k} = e^{-ikX}e^{(-1+ k)\phi}
}
is transformed to
\eqn\etranse{
V^-_{k} = exp\{\lfr1{2\sqrt{p(p+1)}}( [2pk+1]iX +[2pk -(2p+1)]\phi)\}
}
If we put  $k= \lfr{pn + \alpha +1}{2p}$ for $V^{+}_{k} (k>0)$ and
$k= \lfr{-pn -\alpha -1}{2(p+1)}$ for $V^{-}_{k} (k<0)$.
where $n>0$ and $\alpha = 0, \ldots , p-2$,  we obtain all the
DK states of the $c<1$ model.  Though this parametrization\elio\
of the matter and Liouville momenta is somewhat unusual it is perfectly
in accordance with requirements of comparing scaling dimensions that
arise from matrix models.
    This transformation also rotates the ring generators $x$ and $y$ to those
of the $c<1$ model. Thus we can use some of the results of the $c=1$ case
directly in our cases. The ring multiplication trivially follows. So does the
action of the ring element on the DK states. This invariance can easily be
seen if one rewrites the ring elements in light-cone coordinates
($\phi\pm iX$). Now the Lorentz transformation corresponds to
$$\eqalign{
(\phi+iX) \longrightarrow  \gamma (\phi+iX)\quad,\cr
(\phi-iX) \longrightarrow \lfr1\gamma (\phi-iX)\quad,\cr
}
$$
where $\gamma=\sqrt{\lfr{p}{p+1}}$.
The correlation function of N-DK states in $c<1$ is identical to that of
N tachyon
operators in the $c=1$ case if the values of momenta $k$ are chosen
to be as given above. Of course, for these special values $(N,M)$
correlators no longer vanish and one has to extend the calculations of the
$c=1$ case. We can also use the ring elements exactly as in the $c=1$ case
to show that all the momentum dependence of the correlation functions is only
in the form of leg factors\KACH. These issues will be dealt with elsewhere
\GJJ.

\newsec{Matrix model observables in the continuum}
Notwithstanding the interesting structure of the space of states in this
theory, the key question of the identification of the matrix model
observables remains to be answered. We present in this section, a
possible answer to the question, though a complete proof would be
seen to be still lacking. As we shall soon see, these candidates are exact
states but do not appear to decouple in correlation functions.

 The LZ states have, as noted earlier $\beta \leq \beta_0$, with
negative ghost number or $\beta \geq \beta_0$ with positive ghost
number. However the matching of scaling dimensions of the operators with
those of the matrix model requires that $\beta \leq \beta_0$. Since the
Liouville scalar is non-compact, this must be true in both the
holomorphic and non-holomorphic sectors. Clearly if we construct
negative ghost number operators with $\beta \leq \beta_0$ in both
holomorphic and anti-holomorphic sectors, then in any correlation
function the ghost number cannot be balanced without introducing the
positive ghost number operators with $\beta \geq \beta_0$ which of
course will lead to the wrong scaling behaviour with respect to the
matrix model. The question is whether we can introduce operators of
positive ghost number that have $\beta \leq \beta_0$ in some systematic
fashion.

In the construction of LZ states from the DK states described in
\GJJM, it was noticed that in the negative ghost number case, the
LZ condition $\beta<\beta_0$ followed easily. But the same could not
be said for the positive ghost number sector. The whole construction
works for both the possible dressings. But this does not seem to agree
with the cohomology analysis of LZ and BMP. This is resolved when one
notices that for the case of dressing given by $\beta<\beta_0$, the
same states can be constructed by other means which makes them exact.
We shall show this by a simple example. Consider the oscillator state
which replaces the null over the matter vertex operator labelled by
$(1,1)$. This is a level 1 oscillator state. This is given by
\eqn\enull{
\wm11 = \pa X \vm11\quad.
}
The possible Liouville dressings are given by $\beta^>=\beta_{1,1}$ and
$\beta^<=\beta_{-1,-1}$, the former corresponding to $\beta<\beta_0$. The LZ
state at ghost number $1$ is given by,
\eqn\elza{\eqalign{
\ket{W^>}&= \pa X\wm11\otimes\ket{\beta^>}\otimes\gh\quad,\cr
\ket{LZ^>}& = Q_B \ket{W^>}\quad,\cr
        &= c_{-1} \wm11\otimes\ket{\beta^>}\otimes\gh\quad{\rm and}\cr
\ket{W^<}&= \pa X\wm11\otimes\ket{\beta^<}\otimes\gh\quad,\cr
\ket{LZ^<}& = Q_B \ket{W^<}\quad,\cr
        &= c_{-1} \wm11\otimes\ket{\beta^<}\otimes\gh\quad.
}}
The state $\ket{LZ^>}$ is not truly exact since $\ket{W^>}$ is not
closed in the Felder cohomology. The state $\ket{LZ^>}$ would also
seem to be not exact following similar arguments. But, it can be
constructed by another means where it is clearly BRST exact.
\eqn\eexact{\eqalign{
\ket{W_L}&= \pa \phi\wm11\otimes\ket{\beta^<}\otimes\gh\quad,\cr
\ket{LZ^<}&= Q_B \ket{W_L}\quad.
}}
Unlike, $\ket{W^<}$, $\ket{W_L}$ is closed under the action of $Q_F$
and hence the wrong dressing state is BRST exact. However a $Q_F$ map
on both $\ket{W^>}$ and $\ket{W^<}$  gives a DK state that has the
matter charge $\alpha_+$, namely that of one of the matter screening
operators. The choice of wrong Liouville dressing therefore would give us a DK
state that is one of the matter screening operators. In the other case,
we would get an operator with the matter screening charge and the dual
to the identity in the Liouville sector. This Liouville charge however
will create problems in fixing the scaling behaviour
of the correlation function.


Thus we see that at least one of the matter screening operators is
associated with a LZ state that is not in the double cohomology.
{}From the point of view of the DK type of states, the screening
operator plays a crucial role in computing correlation functions.
Hence a correlation function involving such an operator should
presumably be non-vanishing. Thus a correlation function involving a
operator of the form\foot{The bar
refers to the anti-holomorphic sector.}
 ${LZ}^{-n}\otimes{\bar{LZ}}^{-n}$
and a string of operators of the form $c\pa c$ and ${\bar c}{\bar\pa}{\bar c}$
 would preserve ghost number
and would have the same scaling dimensions as a one-point function of
the corresponding DK operator. In general, one would
expect such a correlator to be non-zero. We can see this by generalising the
procedure of the repeated use of the $Q_B$ and $Q_F$ descent that was
outlined in \GJJM for the case of the three-point function.
In the case
of four-point functions and higher we expect various subtleties of the
contact terms of $Q_B$  to come into play; however it is clear that
we could expect it to be non-zero.

This procedure suggests a more general construction of a set of
operators with ghost number zero where the string of $c \partial c$ operators
act effectively at the same point as the operator associated with the
state $\ket{LZ}^{-n}\otimes \ket{\bar{LZ}}^{-n}$. The physical operator in a
symmetric form
therefore could be written as
\eqn\ephys{
	(M-{\bar M})^n (LZ)^{-n}({\bar {LZ}})^{-n}\quad.
}
These operators are obviously closed under $Q_B$  and are very similar to the
zero-momentum dilaton in critical bosonic strings and the operators of
topological gravity. They are also exact from the argument presented earlier
for the exactness of the positive ghost number LZ operators with
$\beta<\beta_0$.

\newsec{Conclusions}
There is still work to be done before the nature of the equivalence between
the continuum formulation and the matrix model and topological formulations
of $c<1$ theories coupled to gravity is fully clear. However it is clear that
the topological theory has a close relation to the underlying Coulomb gas
formalism of the continuum formulation. The extra complication s of screening
etc. make the final picture less clear. The ring structure is particlarly
useful enabling more detailed calculations with DK states. Computations with
the physical operators introduced in the last section are perhaps the most
complicated, as is already evident from the work done on the zero-momentum
dilaton of critical string theory\DOY \TCCT
the analog of the recursion relations of topological models would be
reproduced here with these operators. We have also not isolated the algebras
of $W_n$ type that should appear in these models.

One of the most important problems is the implementation of the
truncation of the "wrong" edge states. We believe that this is the source of
the essentially non-linear algebra of $W_n$ type that appears in these
models. Without this truncation we would see only the essentially linear part
of the algebra.

The formulation of the ring structure here is somewhat different from
the approach of Sarmadi and Kanno \SAR who do not use the Coulomb gas
construction and have a ring of operators of different ghost numbers in
contrast to our approach. It would be interesting to study how these two
different approaches are related. While readying this paper for publication,
we also received their preprint \SARa.

{\leftline {\bf Acknowledgements:}}

We thank Parthasarathi Majumdar for useful discussions and constant
encouragement. S.G. would like to thank H. Sarmadi for useful discussions
as well as the organisers of the Summer Workshop on Strings at the
International Centre for Theoretical Physics, Trieste for the
opportunity to present this work.

\appendix{A}{Cohomology argument for the existence of ring elements}

We do not set out here to explain the full mathematical machinery that is
needed,
especially that of spectral sequences, for the analysis of the double
cohomology. We merely take the results as given in Bouwknegt, McCarthy
and Pilch \BMP. Further mathematical details are given in the book
of Bott and Tu \BT. The double cohomology can be analysed by
either  imposing the $Q_B$ cohomology first or the $Q_F$ cohmology on
 the double complex $(\CF_{m',m}
\bigotimes \CF_{L} \bigotimes \CF_{gh})$, henceforth denoted by $K$.
We shall start with the latter. Note that we wish to include both the
resolution $\CF_{m',m}$ and the dual resolution $\CF_{p'-m', p-m}$.
The $Q_F$ cohomology is non-trivial only for $H^{(0)}_{Q_F}(K)$.
However the dimensionality of this is $2$ since we have included both
the resolutions, viz. we get both the matter momenta $\al_{m',m}$ and
its dual. We can now impose the $Q_B$ cohomology, obtaining
$H^{(n)}_{Q_B}(H^{(0)}_{Q_F}(K))$. The total cohomology for negative
ghost number is given by
\eqn\coha{\eqalign{
H^{(-n)}_D &= \sum_{p+q=-n}H^{p}_{rel}(H^{q}(\CF_{M}\otimes\CF_{L}
\otimes \CF_{gh},Q_F), Q_B) \cr
&=H^{(-n)}_{rel}(H^{(0)}(\CF_{M}\otimes\CF_{L}
\otimes \CF_{gh},Q_F), Q_B)\quad .
}}
It turns out as we explain below that, in fact,
\eqn\cohb{\eqalign{
H^{(-n)}_D &= \sum_{p+q=-n}H^{q}(H^{p}_{rel}(\CF_{M}\otimes\CF_{L}
\otimes \CF_{gh},Q_B), Q_F) \cr
&= H^{-n+1}(H^{-1}_{rel}(\CF_{M}\otimes\CF_{L}
\otimes \CF_{gh},Q_B), Q_F) \cr
&\phantom{=}+H^{-n}(H^{0}_{rel}(\CF_{M}\otimes\CF_{L}
\otimes \CF_{gh},Q_B), Q_F)\quad .}}
Since $H^{(-n)}_{D}$ computed both ways should be the same, and the DK
states exhaust  the elements of $H^{-n}_{rel} (
H^{(0)}(\CF_{M}\otimes\CF_{L}
\otimes \CF_{gh},Q_B), Q_F)$, the ring elements are clearly identified
with the rest.

Before deriving \cohb, we first illustrate the technique of spectral sequences
for filtered complexes by deriving \coha. For the double cohomology in
question, we have one spectral
sequence $\{E_{r} , d_r\}$ where $E_r$ is bigraded  and
\eqn\cohc{\eqalign{
E_{1}^{p,q} &= H_{Q_F}^{p,q}(K)\quad, \cr
E_2^{p,q} &= H_{Q_B}^{p,q}H_{Q_F}(K)\quad,\cr
&\vdots\phantom{junk}
}}
and $d_r:E_{r}^{p,q}\rightarrow E_{r}^{p+r,q-r+1}.$ Note that in this
notation, $H_{Q_F}^{p,q}$ is the $p$th $Q_F$ cohomology in the $q$th
complex and $H_{Q_B}^{p,q}$ is the $q$th $Q_B$ cohomology in the $p$th
complex. Obviously, $p$ is the tower number and $q$ is the ghost number
in the complex $K$. Since $H_{Q_F}$ is non-zero only for $p=0$, we have
$$
E_{2}^{0,n}=H_{Q_B}^{0,n}H_{Q_F}(K)
$$
It is clear that $d_2$ is zero as there is no element in $E_2^{2,n-1}$.
Hence $E_2=E_3=\ldots=E_{\infty}$ and
$H_D^{(n)}=H_{Q_B}^{0,n}H_{Q_F}(K)$.
This is indeed \coha .
We can also construct another spectral sequence, $\{E_r^{\prime} ,
d^{\prime}_r\}$,
where $E^{\prime}_r$ is bigraded and
\eqn\cohd{\eqalign{
E_1^{\prime p,q} &= H_{Q_B}^{p,q}(K)\cr
E_2^{\prime p,q} &= H_{Q_F}^{p,q}H_{Q_B}(K)\cr
&\vdots\phantom{junk}
}}
and $d_r^{\prime} : E_r^{\prime p,q} \rightarrow E_r^{\prime p-r+1, q+r}.$
If we are in the sector of Liouville momenta with $\beta < \beta_0$,
we get non-zero elements only in $E_2^{\prime(n,0)}=
H_{Q_F}^{n,0}H_{Q_B}(K)$ and  in $E_2^{\prime (n-1,1)}=
H_{Q_F}^{n-1,1}H_{Q_B}(K)$.
Hence again $d_2^{\prime}$ is zero and
$E_2^{\prime}=E_3^{\prime}=\ldots=E_{\infty}^{\prime}$.
Thus
\eqn\cohe{\eqalign{
H^{(n)}_D(K) &= \oplus_{p+q=n} E^{(p,q)}_{\infty}(K) \cr
             &= \oplus_{p+q=n}E^{(p,q)}_2(K)\cr
             &= H_{Q_F}^{(n-1,1)}H_{Q_B}(K) \oplus
H_{Q_F}^{(n,0)}H_{Q_B}(K).
}}
One can always rotate from one quadrant to another in the
two gradings and hence we have \coha and \cohb.

\listrefs
\bye